\documentclass[]{jfm}

\usepackage{graphicx,float}

\usepackage{amsmath}%
\usepackage{subfig}
\usepackage{psfrag}
\usepackage[usenames]{color}
\usepackage{float}
\usepackage{natbib}
\usepackage{pifont}
\usepackage{relsize}
\usepackage{graphicx}

\def\drawline#1#2{\raise 2.5pt\vbox{\hrule width #1pt height #2pt}}

\def\drawline#1#2{\raise 2.5pt\vbox{\hrule width #1pt height #2pt}}

\def\trian{\raise 1.25pt\hbox{$\scriptscriptstyle\triangle$}\nobreak}

\def\square{${\vcenter{\hrule height .4pt
        \hbox{\vrule width .4pt height 3pt \kern 3pt
        \vrule width .4pt}
        \hrule height .4pt}}$\nobreak\ }

\title[]{Flow dynamics and wall-pressure signature in high-Reynolds number overexpanded nozzle with free shock separation}

\author[E. Martelli, L. Saccoccio, P.P. Ciottoli, C.E. Tinney, W.J. Baars, M. Bernardini]
{E. \ns Martelli$^{1}$,
\ns L. \ns Saccoccio$^{2}$,
\ns P.P. \ns Ciottoli$^{2}$,
\ns C.E. \ns Tinney$^{3}$
\ns W.J. \ns Baars$^{4}$
and
\ns M. \ns Bernardini$^{2}$
}

\affiliation{
$^1$Dipartimento di Ingegneria,
Universit\`a degli Studi della Campania  "L. Vanvitelli", \break Caserta,
81100, Italy \\[\affilskip]
$^2$Dipartimento di Ingegneria Meccanica e Aerospaziale, 
Sapienza Universit\`a  di Roma\break Roma, 00184, Italy \\[\affilskip]
$^3$Department of Mechanical Engineering,
University of Texas at Austin, \break
Austin, TX 78713, USA \\[\affilskip]
$^4$School of Engineering,
RMIT University,  \break
Melbourne, Victoria 3010, Australia \\[\affilskip]
}

\date{\today}
\begin{document}
\maketitle

\begin{abstract}
A delayed detached eddy simulation (DDES) of an overexpanded nozzle flow with shock-induced separation
is carried out at a Reynolds number, based on the nozzle throat diameter and the stagnation chamber
properties, equal to $1.7\times 10^7$.
The flow unsteadiness, characterised by self-sustained shock oscillations, induces
local unsteady loads on the nozzle wall as well as global off-axis forces.
Despite several studies in the last decades, a clear physical understanding
of the driving factors of the unsteadiness is still lacking.
The geometry under investigation is a sub-scale truncated ideal contour (TIC) nozzle,
which has been experimentally tested at the University of Texas at Austin.
Under the current conditions, the nozzle operates
in a highly-overexpanded regime (relevant to  the nozzle ignition) and
comprises a conical separation shock within the nozzle contour, merging
into a Mach disk in the nozzle centre.
Our current study focuses on the unsteady pressure signature on the nozzle wall,
through the use of Fourier-based spectral analysis performed in time and in the
azimuthal wavenumber space. The numerical data from the
detached eddy simulation well agrees with the experimental measurements in terms of
mean and fluctuating wall pressure statistics.
The frequency spectra are characterised by the presence of a large bump in
the low frequency range associated to a breathing motion of the shock system
and a broad and high amplitude peak at high frequencies generated by the turbulent activity
of the detached shear layer.
Moreover, a distinct peak at an intermediate frequency (of the order 1000 Hz)
is observed to persist in the wall-pressure spectra along the nozzle wall.
The analysis of the pressure signals in the azimuthal wavenumber space indicates that this peak is clearly associated to
 the first (non-symmetrical) pressure mode ($m = 1$) and it is thus connected to the generation of side loads.
Furthermore, it is found that the unsteady Mach disk is characterised by an intense
vortex shedding activity and the interaction of these
vortices with the second shock cell is a key factor in
the sustainment of an aeroacoustic feedback loop within the nozzle.
\end{abstract}

\section{Introduction}
\label{sec:intro}
\begin{figure}
 \centering
 \includegraphics[width = 0.97\textwidth]{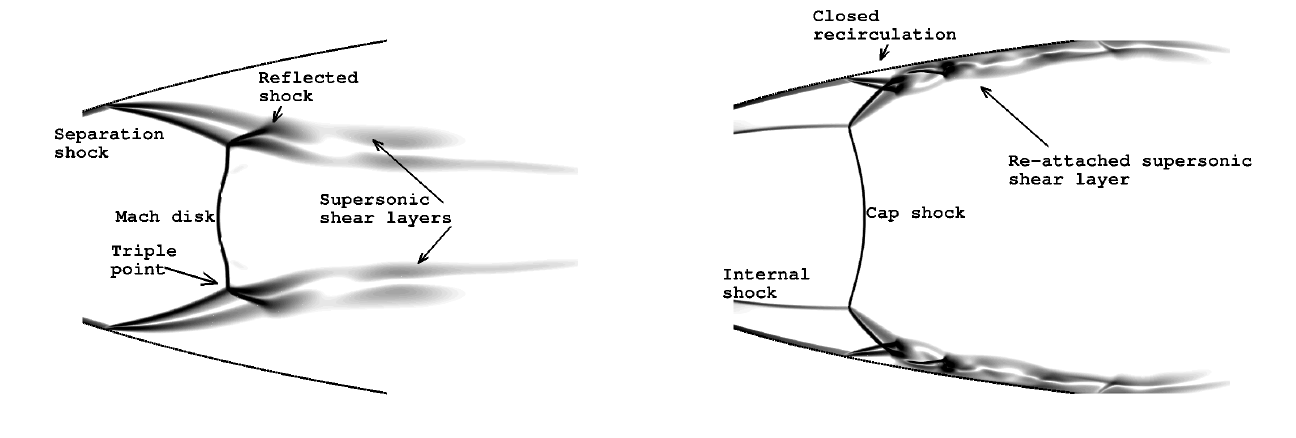}
 \caption{Schematic of the two typical flow separation regimes in overexpanded rocket nozzles:
          left) free shock separation (FSS) in a TIC nozzle; right) restricted shock separation (RSS) in a TOP nozzle.}
 \label{fig:gradrho_2d}
\end{figure}

The performance of first-stage liquid rocket engines is highly dependent on the fluid dynamic behaviour of the expansion nozzle and for optimisation purposes, 
large values of the ratio between the exit and throat areas are desirable. 
The maximum limit to this ratio is imposed by the need to avoid internal flow separation, since at sea level 
the flow is highly overexpanded. 
However, during the start-up phase the chamber pressure is below 
the design pressure and the flow separates from the nozzle wall. This condition is characterised by complex physical features, including
the formation of a shock-wave system that adapts the exhaust flow to the higher ambient pressure, 
shock-wave/boundary-layer interactions, and a turbulent recirculating zone with a pulsating pressure field. As a global effect, the nozzle experiences 
non-axial forces, known as side-loads,
which can be of sufficient strength to cause structural damage to the engine~\citep{nave}.
Several investigations~\citep{chen94,nasuti98,hagemann2000,ostlund02} demonstrated the 
existence of two kinds of flow separation regimes: the free shock separation (FSS) and 
the restricted shock separation (RSS). A brief and complete literature review can be found in \citet{hadjadj09}. 
The occurrence of a specific pattern (FSS or RSS) is mainly determined by two parameters, the geometry and the nozzle pressure ratio (NPR), i.e. the ratio between the chamber and the ambient pressure. Truncated ideal contour (TIC) and conical nozzles  
display only the FSS type, which is shown in figure~\ref{fig:gradrho_2d} (left panel). 
The adverse pressure gradient causes the separation of the flow at a certain axial location and the consequent formation of compression waves which 
coalesce in a conical shock (separation shock). This shock reflects on the nozzle axis of symmetry through a Mach disk and a second oblique leg (reflected shock). These three shocks meet at the 
so-called triple point (TP). This type of interaction is named {\it free} since the
separated shear layer never reattaches to the nozzle wall.
Thrust-optimised contour (TOC) and thrust-optimised parabolic (TOP) nozzles 
display both FSS and RSS depending on the nozzle pressure ratio. Figure~\ref{fig:gradrho_2d} (right panel) shows the restricted shock separation, 
which is characterised by a cap-shock pattern~\citep{nave} and the re-attachment of the 
separated shear layer to the wall with a huge recirculation bubble (not shown in the figure) on the nozzle centreline~\citep{nasuti98}.

Since the early investigations reported in literature, it was argued that lateral forces are mainly due to oscillations of the internal shock system inside the nozzle. In TOP and TOC nozzles, it has been observed that the side-load history displays an important peak
when the flow transitions from FSS to RSS~\citep{frey2000,ostlund02}.
The first research efforts were directed towards the development of analytical and empirical methods to predict the occurrence of side-loads and quantify their magnitude.
With this perspective, \citet{schmucker} developed a model based on the idea of a tilted separation line, while \citet{Dumnov1996} considered oscillations of the separation line excited by random pressure pulsations in the separated flow region.
These methods rely on many approximations and are primarily tailored to design purposes.
Only recently, research has been directed towards a more physical understanding of the origin of aerodynamic loads and the consequent generation of lateral forces. Our present work focuses on the FSS pattern, as it is common to all kind of nozzles and is characterised by an intense side-load activity~\citep{ruf2010}.
\citet{Baars2012} and \citet{baars:2013} have carried out an experimental campaign to analyse the unsteadiness of the wall pressure in a cold-gas, sub-scale parabolic nozzle in both FSS and RSS configurations. They found that the power spectral density of the wall pressure fluctuations in FSS state is characterised by low and high-frequency broad peaks, associated to the shock-wave/boundary-layer interaction and to the development of the turbulent shear layer.
Furthermore, the authors were able to isolate the effect of the anti-symmetric azimuthal mode, the only one linked to the side-load generation.
More recently, \citet{Jaunet2017} have observed experimentally the evolution of the FSS pattern and the associated azimuthal Fourier modes by varying the NPR in a sub-scale TIC nozzle. As a major result, they showed that the low-frequency mode is mostly axi-symmetric and confined in the nozzle, while the developing turbulent shear layer leaves a high-frequency signature in the pressure and velocity fields, both inside and outside the nozzle. In addition, they also found highly organised pressure structures at an intermediate frequency range, mainly associated with the anti-symmetric pressure mode. They argued that these structures may be attributed to a screech-like mechanism~\citep{Raman99} rather than to a transonic resonance~\citep{wong:05jpp}.

All the experimental work done so far seems to confirm that the various 
frequency modes identified are general features of the free shock separation pattern. However, the experiments on axi-symmetric nozzles
suffer from the lack of flow measurements inside the nozzle itself,
due to the challenging flow conditions and absence of optical access.
Therefore, numerical simulations represent an important complementary tool to gain a more complete  insight into the flow physics of
separated rocket nozzle flows,
and provide the possibility of addressing important open questions:
i) which kind of dynamics and frequency content characterise the vortices in the initial part of the supersonic shear layer? 
ii) If a screech-like mechanism is present due to the interaction of 
the turbulent shear layer and the shock cell inside the jet, which role do the Mach disk and the subsonic flow region behind it play in the feedback loop? 

Unsteady Reynolds averaged Navier-Stokes (URANS) equations have been used in the past to evaluate the level of the side-loads in sub-scale models and rather good results were obtained 
by~\citet{Deck2002} and~\citet{Deck2004}. Nevertheless, modelling the global effect of the turbulent scales as done in the URANS approach could hide the important flow processes leading 
to the formation of the aerodynamic unsteady loads.
On the other hand, performing a wall-resolved large eddy simulation (LES) of this kind of flows, characterised by Reynolds numbers of the order of $10^7$,
requires impractically high computational expenses. A valid alternative is then represented by the use of the detached eddy simulation~\citep{spalart97} (DES), a hybrid RANS/LES method that allows to simulate high-Reynolds number flows involving massive flow separation.
In fact, in this approach, attached boundary layers are treated in RANS mode, lowering the computational requirements, while the most energetic turbulent scales of  
separated shear layers and turbulent recirculating zones are directly described by the LES mode
of the method.
Nevertheless, very few DES simulations of separated nozzle flows can be found in literature. \citet{Deck2009} and \citet{Shams2013} presented a delayed detached eddy simulation (DDES) of the end-effect regime in an axi-symmetric nozzle flow characterised by a RSS  pattern. 
While the  main flow properties were rather well reproduced by the simulations, the computed main frequency resulted to be higher than in the experiment.
As far as FSS is concerned, to the authors knowledge, the only DES available has been reported by
\citet{Larusson2013}, who exclusively focused on the prediction of the side-load magnitude.

In this study, we present the results of a detached eddy simulation of a TIC nozzle with flow separation. The geometry has been tested at the University of Texas at Austin, where
measurements of the unsteady wall-pressure signals were recorded. The analysis is first focused on the behaviour of the wall-pressure signature and on the physical mechanisms leading to the generation of the aerodynamic loads. The numerical results are then used to investigate and characterise those parts of the flow not accessible in the experiments, as the Mach disk region and the initial part of the annular supersonic shear layer.

The paper is organised as follows. First, the experimental apparatus and instrumentation are described
in section~\ref{sec:exp}. Then, the delayed detached eddy simulation approach, the numerical solver and the computational
setup are presented in section~\ref{sec:comp}. After providing an overview of the flowfield organisation in section~\ref{sec:flow_org},
the main features of the wall pressure signature are analysed in section~\ref{sec:wallp}
by means of spectral analysis (through a wavenumber-frequency decomposition) and
the evaluation of space-time correlations. In section~\ref{sec:shear_machdisk} the attention is then focused
on the dynamics of the annular supersonic shear layer and of the Mach disk region,
and a discussion is presented about the possible existence of an aeroacoustic feedback loop within the nozzle.
Conclusions are finally given in section~\ref{sec:conclusions}.

\section{Main features of the experimental campaign}
\label{sec:exp}
\subsection{Facility and nozzle test article}
The rocket-exhaust nozzle under investigation was experimentally tested at The University of Texas at Austin. A short description of the facility is here provided \citep[further details can be found elsewhere,][]{baars:2013,baars:2014a}. An anechoic chamber with interior dimensions of approximately 5.74\,m in length (parallel to the nozzle's centreline), 4.52\,m in width and 3.66\,m in height forms the outer perimeter of the test space. The chamber's wall-design yields a fully-anechoic environment for frequencies larger than 100\,Hz, and allows for the aeroacoustic investigations of rocket-nozzle flows \citep[e.g.][]{donald:2014a}. A rig situated within the centre of the chamber provides the air supply to the nozzle via an instrumented, blown-down compressed air system. The nozzle exhausts through a 1.8\,m$^2$ acoustically-treated duct, while on the upstream side a 1.5\,m$^2$ inlet provides an ambient air supply to the chamber.

\begin{figure}
 \centering
 \includegraphics[width = 0.97\textwidth]{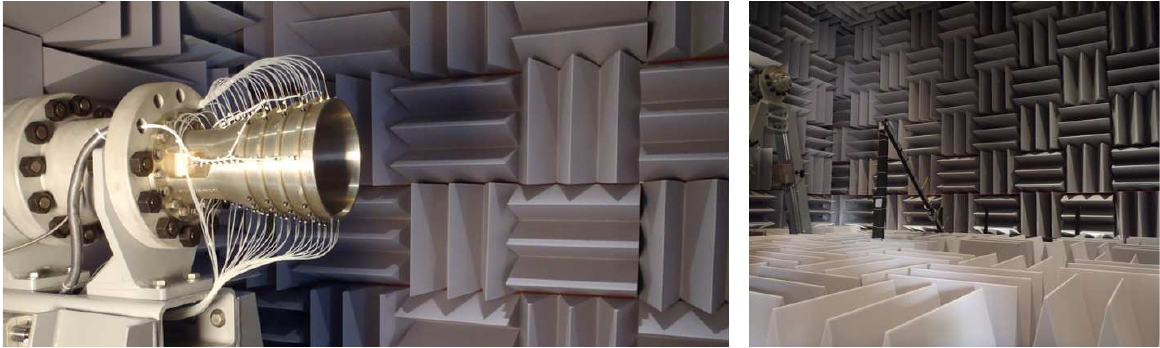}
 \caption{(\emph{left}) Photo of a sub-scale, large-area ratio rocket nozzle installed on the test rig at UT Austin (similar size as the TIC-contoured). (\emph{b}) A view of the anechoic test environment in which the experiments were performed; the nozzle test rig is visible on the left-hand-side.}
    \label{fig:photos}
\end{figure}

The reduced-scale nozzle under investigation is a TIC nozzle, designed by the Nozzle Test Facility team at NASA Marshall Space Flight Center. The length of the diverging nozzle contour is 79\% of that of a 15$^{\circ}$ conical nozzle with the same area ratio; this is a standard truncation length for ideal contour nozzles \citep{rao:1958}. Basic geometric properties of the nozzle are a throat radius of $r_t = 19.0$\,mm, an exit radius of $r_e = 117.43$\,mm and a throat-to-exit length of $L = 350.52$\,mm. The exit-to-throat area-ratio is $A_e/A^* = 38$ and governs a design exit Mach number of $M_e = 5.58$ at a nozzle pressure ratio (NPR) of 970. Photos of the nozzle and the test environment are shown in Figure~\ref{fig:photos}.

\subsection{Operating conditions and instrumentation}
\label{subsec:oci}
\begin{figure}
	\includegraphics[width = 0.97\textwidth]{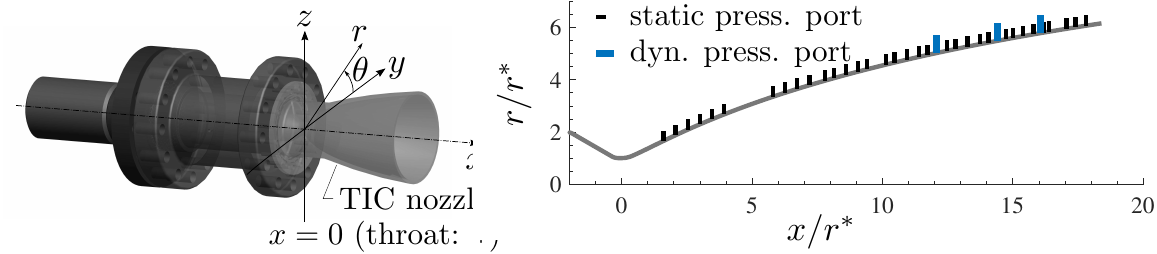}
 \caption{(\emph{a}) Schematic of the nozzle with coordinate system. (\emph{b}) TIC nozzle contour with an indication of the axial locations of the static and dynamic pressure ports.}
 \label{fig:TICsetup}
\end{figure}

For this current study, a constant NPR of nominally 30.35 was maintained for approximately 10 seconds. Data of static and dynamic wall pressure sensors were acquired and form the validation data for the DDES of the flow. Static wall pressures were measured using two Scanivalve DSA3218 gas pressure scanners, to which a total of 34 static pressure ports were connected. These ports were oriented along the axial direction of the nozzle contour, between $x/r = 1.60$ and 17.80. Via this set-up, it is ensured that axial static wall-pressure profiles provide the location of the separation shock foot. For the fluctuating wall pressure, a total of nine time traces are utilized. These time histories were acquired using Kulite XT-140 dynamic pressure transducers with a dynamic range of 100\,psia ($\pm 0.1$\,\% full-scale output), installed in such a way that their protective B-type screens, with a 2.62\,mm outside diameter, were flush with the interior surface. These screens limit the effective frequency response up to 20\,kHz. All channels were sampled simultaneously at a rate of $f_s = 40$ kHz using the appropriate signal conditioning on a National Instruments PXI based system. Transducers were installed at three axial stations in the nozzle: $x/r = 12.07$, 14.40 and 16.07, with three transducers per axial location at different azimuth angles. Finally, Figure~\ref{fig:TICsetup} displays a schematic of the nozzle, the nozzle contour and indicates the locations of the pressure ports.

\section{Computational strategy}
\label{sec:comp}

\subsection{Physical model}

We solve the three-dimensional Navier-Stokes equations for a compressible, viscous, heat-conducting gas
\begin{equation}
 \left .
 \begin{aligned}
   \frac{\partial \rho}{\partial t} + \frac{\partial (\rho \, u_j)}{\partial x_j}
     & = 0, \\
   \frac{\partial (\rho \, u_i)}{\partial t} + \frac{\partial (\rho \, u_i u_j)}{\partial x_j} +
       \frac{\partial p}{\partial x_i} - \frac{\partial \tau_{ij}}{\partial x_j} & = 0, \\
   \frac{\partial (\rho \, E)}{\partial t} + \frac{\partial (\rho \, E u_j + p u_j)}{\partial x_j}
     - \frac{\partial (\tau_{ij} u_i - q_j)}{\partial x_j} & = 0, \label{eq:ns}
 \end{aligned}
 \right .
\end{equation}
where $\rho$ is the density, $u_i$ is the velocity component in the $i$-th coordinate direction ($i=1,2,3$),
$E$ is the total energy per unit mass, $p$ is the thermodynamic pressure.
The total stress tensor $\tau_{ij}$ is the sum of the viscous and the Reynolds stress tensor, 
\begin{equation}
 \label{eq:stress}
 \tau_{ij} = 2 \, \rho \left ( \nu + \nu_t \right ) S^*_{ij} \qquad S^*_{ij} = S_{ij} - \frac 13 \, S_{kk} \, \delta_{ij},
\end{equation}
where the Boussinesq hypothesis is applied through the introduction of the eddy viscosity $\nu_t$, $S_{ij}$ is the strain-rate tensor and $\nu$ the kinematic viscosity, depending on temperature $T$ through Sutherland's law.
Similarly, the total heat flux $q_j$ is the sum of a molecular and a turbulent contribution
\begin{equation}
 q_j = -\rho \, c_p \left ( \frac{\nu}{\mathrm{Pr}} + \frac{\nu_t}{\mathrm{Pr}_t} \right ) \frac{\partial T}{\partial x_j},
\end{equation}
where $\mathrm{Pr}$ and $\mathrm{Pr}_t$ are the molecular and turbulent Prandtl numbers, assumed to be
0.72 and 0.9, respectively.

\subsection{Turbulence modelling}

Because of the high-Reynolds number of the flow investigated in this work, the adopted numerical methodology is the delayed detached eddy simulation~\citep{spalart06}, which belongs to the family of the hybrid RANS/LES methods.
Our implementation is based on the Spalart-Allmaras (SA) turbulence model, which solves a transport equation for a pseudo eddy viscosity $\tilde{\nu}$ 
\begin{equation} 
 \label{eq:sa}
 \frac{\partial (\rho \tilde{\nu})}{\partial t} + \frac{\partial (\rho \, \tilde{\nu} \,u_j)}{\partial x_j} = 
 c_{b1} \tilde{S} \rho \tilde{\nu} +
 \frac{1}{\sigma}
 \left [
  \frac{\partial}{\partial x_j} \left [ \left ( \rho \nu + \rho \tilde {\nu} \right ) \frac{\partial \tilde{\nu}}{\partial x_j} \right ] +
  c_{b2} \, \rho \left ( \frac{\partial \tilde{\nu}}{\partial x_j} \right )^2
 \right ]
  -c_{w1}  f_w \rho \left ( \frac{\tilde{\nu}}{\tilde{d}} \right )^2,
\end{equation}
where $\tilde{d}$ is the model length scale, $f_w$ is a near-wall damping function, $\tilde{S}$
a modified vorticity magnitude, and $\sigma,~c_{b1},~c_{b2},~c_{w1}$ model constants.
The eddy viscosity in Eq.~\ref{eq:stress} is related
to $\tilde{\nu}$ through $\nu_t = \tilde{\nu} \, f_{v1}$, where $f_{v1}$ is a
correction function designed to guarantee the correct boundary-layer behaviour in the near-wall region.
In DDES the destruction term in Eq.~\ref{eq:sa} is designed in such a way that the model reduces to pure RANS in attached boundary layers and to a LES sub-grid scale model in flow regions detached from walls.
This goal is accomplished by defining the length-scale $\tilde{d}$ as
\begin{equation}
 \tilde{d} = d_w - f_d \, \textrm{max} \left (0, d_w-C_{DES} \, \Delta \right),
\end{equation}
where $d_w$ is the distance from the nearest wall, $\Delta$ is the subgrid length-scale that controls the wavelengths resolved in LES mode and $C_{DES}$ is a calibration constant equal to 0.20. 
The function $f_d$, designed to be $0$ in boundary layers and $1$ in LES regions, is defined as
\begin{equation}
 f_d = 1-\tanh{\left [ \left ( 16 r_d \right )^3 \right ]}, \qquad r_d = \frac{\tilde{\nu}}{k^2 \, d_w^2 \, \sqrt{U_{i,j} U_{i,j}}},
\end{equation}
where $U_{i,j}$ is the velocity gradient and $k$ the von Karman constant.
The introduction of $f_d$ distinguishes DDES from the original DES approach~\citep{spalart97} (denoted as DES97), guaranteeing that boundary layers are treated in RANS mode also in the presence of particularly fine grids, for which the spacing in the wall-parallel directions does not exceed the boundary layer thickness. This precaution is needed to prevent the phenomenon of modeled stress depletion, consisting in the excessive reduction of the eddy viscosity in the region of switch (grey area) between RANS and LES, which in turn can lead to grid-induced separation. 

Differently from the original DDES formulation, the sub-grid length scale in this work is defined according to~\citet{Deck2012}, and it depends on the flow itself, through $f_d$ as
\begin{equation}
 \label{eq:delta}
 \Delta = \frac{1}{2} 
 \left [
 \left (1+\frac{f_d-f_{d0}}{|f_d-f_{d0}|} \right ) \, \Delta_{\textrm{max}} +
 \left (1-\frac{f_d-f_{d0}}{|f_d-f_{d0}|} \right ) \, \Delta_{\textrm{vol}}
 \right ],
\end{equation}
where $f_{d0} = 0.8$, $\Delta_{\mathrm{max}} = \max (\Delta x, \Delta y, \Delta z)$ and $\Delta_{\mathrm{vol}} = (\Delta x \cdot \Delta y \cdot \Delta z)^{1/3}$.
The main idea of this formulation is to take advantage of the $f_d$ function to switch
between $\Delta_{max}$, needed to shield the boundary layer, and $\Delta_{\mathrm{vol}}$,
needed to ensure a rapid destruction of modelled viscosity to unlock the Kelvin-Helmholtz instability
and accelerate the passage to resolved turbulence in the separated shear layer.
The problem of modeled stress depletion,
the need of avoiding the delay in the onset of shear instabilities and, more in general,
the management of the hybridization strategy of RANS and LES are
well-known challenges, whose solutions are still today
the subject of modelling efforts~\citep{moser18}.

\subsection{Flow solver description}

The simulations have been carried out by means of an in-house compressible flow solver,
which solves the compressible Navier-Stokes equations on structured grids.
In the flow regions away from the shock, the spatial discretization consists of a centred, second-order, finite volume scheme~\citep{Pirozzoli2011}. The approach is based on an energy consistent
formulation that makes the numerical method extremely robust without the addition of numerical dissipation~\citep{Pirozzoli2011}.
This feature is particularly useful in the flow regions treated in LES mode, where in addition to the molecular, the only relevant viscosity should be that provided by the turbulence model.
Near discontinuities, identified by the Ducros shock sensor~\citep{ducrosetal99},
the scheme switches to third-order Weno reconstructions for
cell-faces flow variables. The gradients normal to the cell faces needed for the viscous fluxes,
are evaluated through second-order central-difference approximations,
obtaining compact stencils and avoiding numerical odd-even decoupling phenomena.
A low-storage, third-order Runge-Kutta algorithm~\citep{Bernardini2009} is used for
time advancement of the semi-discretized ODEs' system.
The code is written in Fortran 90, it uses domain decomposition and it fully exploits the message passing interface (MPI) paradigm for the parallelism.
The accuracy and reliability of the flow solver have been assessed in a series of recent studies
dedicated to DDES of shock waves boundary layer interactions~\citep{martelli2017,martelli2019,memmolo2018}.

\subsection{Test case description and computational setup}

\begin{figure}
 \centering
	\includegraphics[width = 0.97\textwidth]{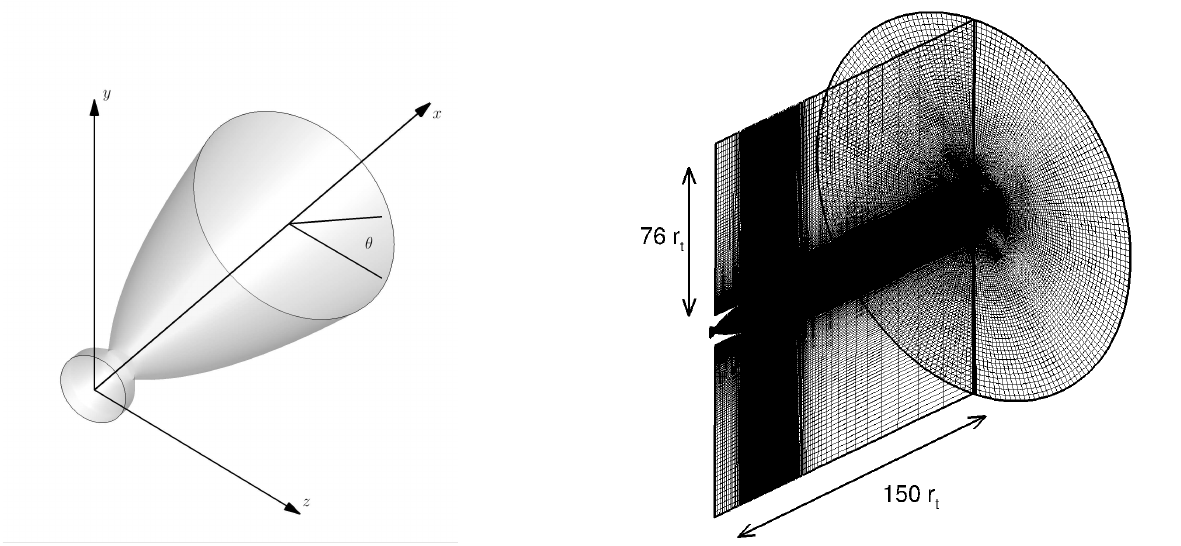}
 \caption{Schematic of (left) the reference coordinate system and (right) of the computational mesh adopted for DDES.}
 \label{fig:mesh}
\end{figure}
The parameters of the simulation were selected to reproduce the operating conditions described in~\ref{subsec:oci}, characterised by total pressure $p_0 = 3.035$ MPa and ambient pressure $p_a = 0.1$ MPa, resulting in a nozzle pressure ratio NPR = 30.35.
The total temperature $T_0$  and the static ambient temperature $T_a$ have been set equal to 300 K. The nozzle Reynolds number, evaluated assuming the throat radius as reference 
length and density $\rho_0$, speed of sound $a_0$ and molecular viscosity taken at the stagnation-chamber condition $\mu_0 = \mu (T_0)$ is
$$Re = \frac{\rho_0 a_0 r_t}{\mu_0} = \frac{\sqrt{\gamma}}{\mu_0}\frac{p_0 r_t}{\sqrt{R_{air} T_0}}=1.75 \cdot 10^7.$$
The three-dimensional computational domain has been designed to include the nozzle and an extended portion of the external ambient, as shown by the schematic in figure~\ref{fig:mesh}. Starting from the throat station, the outflow boundary extends up to 150 throat radii in the longitudinal direction, while in the radial direction the external boundary is placed at 76 $r_t$ from the symmetry axis. As far as the boundary conditions are concerned, total pressure, total temperature and flow direction are imposed at the nozzle inflow. An assigned downstream pressure equal to $p_a$ is prescribed on the outside boundaries, except for the outflow boundary on the right of the computational domain, where non-reflecting boundary conditions are imposed.
The nozzle walls are treated by prescribing the no-slip adiabatic condition.

To select the mesh resolution, a preliminary sensitivity study was performed by means of steady-state axi-symmetric RANS simulations.
The final grid includes approximately 85 million cells, with 192 cells in the azimuthal direction (see figure~\ref{fig:mesh}).
The RANS solution was used to initialise the three-dimensional DDES computation.
To promote the development of turbulent structures and the passage from modelled
to resolved turbulence, random perturbations were added at the initial time
of the 3D simulation to the streamwise velocity field,
with a maximum magnitude of $3\%$ of the inflow velocity.

The computation was run with a time step $\Delta t = 6.2 \cdot 10^{-8}$s and a
relatively long time span was simulated $T = 0.083$ s, which
guarantees coverage of frequencies down to at least $f_{min} \approx 12 $ Hz.
A total of 500 full three-dimensional fields have been collected at time intervals of
$1.66 \cdot 10^{-4}$ s for post-processing purposes. Furthermore, samples of the pressure field at the wall
and in an azimuthal plane have been recorded at shorter time intervals of
$3.1 \cdot 10^{-6}$ s to guarantee sufficient resolution for the frequency
analysis. Running with 2304 processors of the Tier-0 system Marconi (Cineca supercomputing facility),
the cost of the simulation was approximately 3.17 Mio CPU hours. 

\section{Flowfield organisation}
\label{sec:flow_org}

\begin{figure}
 \centering
	\includegraphics[width = 0.97\textwidth]{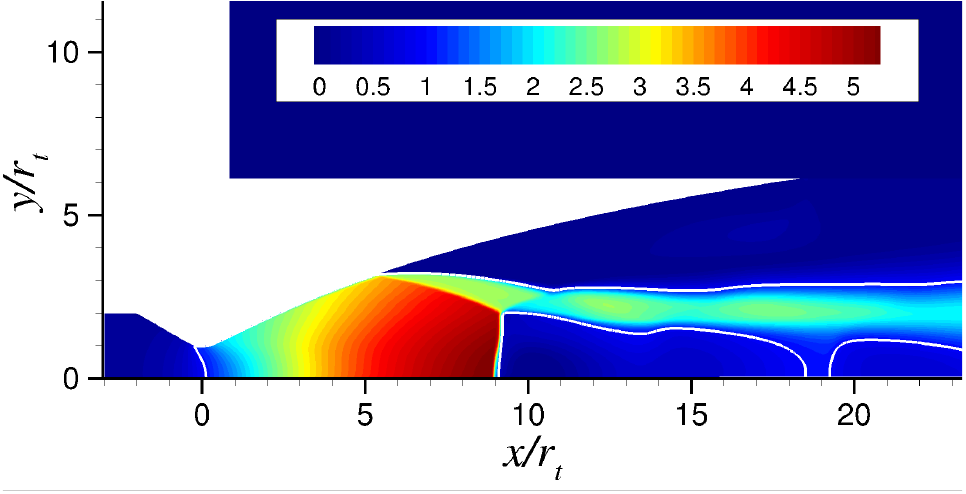}
 \caption{Contours of the averaged Mach number field from DDES. The white solid line denotes the sonic level.}
 \label{fig:mach}
\end{figure}
The salient features of the free shock separation pattern are illustrated in figure ~\ref{fig:mach}, where contours of the mean Mach number field obtained by averaging in the azimuthal direction and time are reported in the x-r plane.
Due to the flow overexpansion a shock system appears inside the nozzle to adapt the exhaust pressure to the ambient pressure. The consequent strong adverse pressure gradient induces the flow separation and the formation of a shear layer that does not reattach to the wall. The wall region is dominated by an important subsonic recirculating flow.
The shock system comprises a conical shock, which is reflected as a Mach disk on the nozzle axis. The reflection is completed by a second conical shock, which deflects the inclined supersonic annular jet in a direction nearly parallel to the nozzle axis. Behind the Mach disk the flow is subsonic at the beginning, then it expands and, through a 
fluid-dynamic throat, 
again accelerates up to a supersonic velocity, requiring the occurrence of a new shock to adapt the jet pressure to the ambient level.
It is worth to point out that the flow pattern resembles that of a classical overexpanded jet but, differently from aeronautical applications, the jet starts well inside the nozzle due to flow separation and it is then a wall-confined jet.

\begin{figure}
 \centering
	\includegraphics[width = 0.7\textwidth]{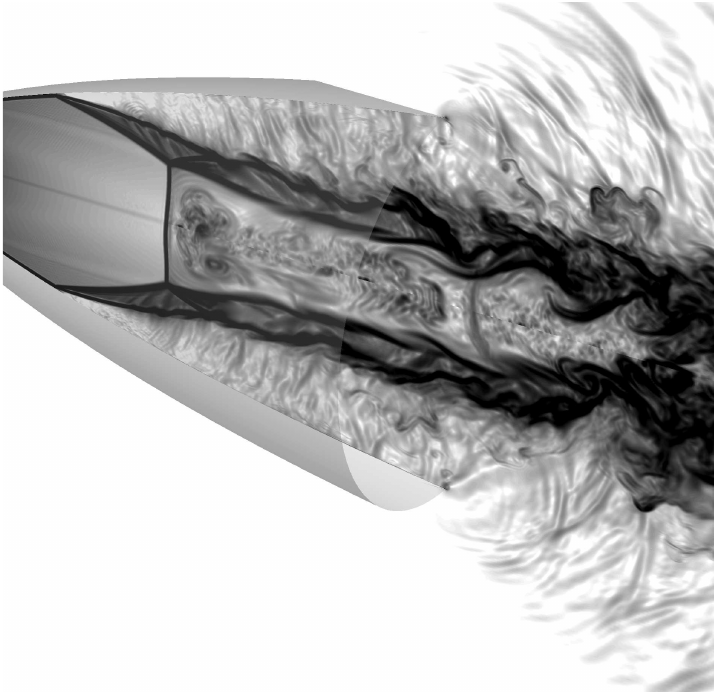}
 \caption{Visualisation of the instantaneous density-gradient magnitude (numerical Schlieren) in an azimuthal plane.}
 \label{fig:grad}
\end{figure}

\begin{figure}
 \centering
	\includegraphics[width = 0.7\textwidth]{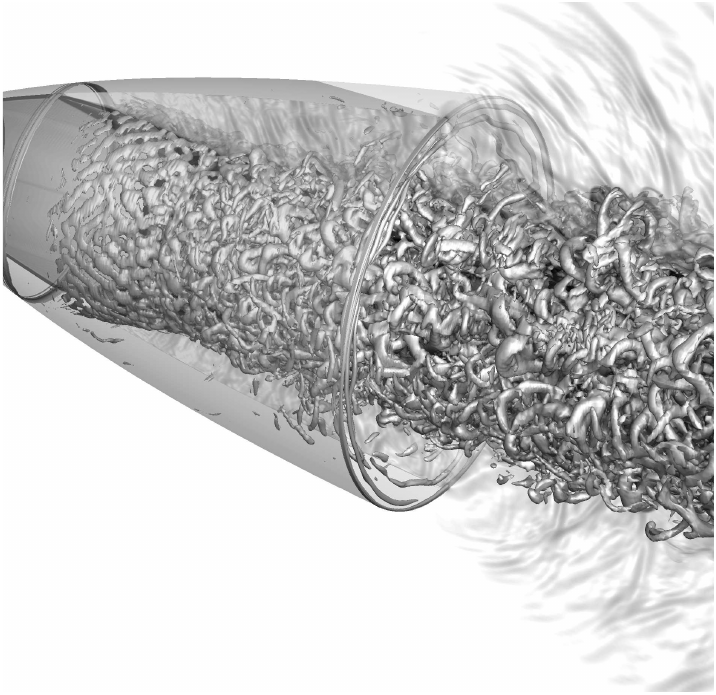}
 \caption{Visualisation of turbulent structures through an iso-surface of the $Q^*$ criterion.}
 \label{fig:Q}
\end{figure}

The unsteady features of the flow are highlighted by showing contours of the density-gradient magnitude in figure~\ref{fig:grad} and an iso-surface of the Q-criterion~\citep{hunt88} in figure~\ref{fig:Q}. The latter is a well-known qualitative method used to identify tube-like vortical structures and in this analysis it has been modified to account for the effect of compressibility~\citep{pirozzoli2008}. Let $A=\nabla{\bf u}$ be the gradient velocity tensor and $A^*=(A-\frac{1}{3}\nabla \cdot {\bf u} I)$ its traceless part, turbulent structures are extracted by visualising regions with a positive iso-value of the second invariant of $A^*$, defined as $Q^*=-\frac{1}{2} A^*_{ij}A^*_{ji}$, since in these regions rotation exceeds the strain. 
Various features are clearly visible in the visualisation of figures~\ref{fig:grad} and \ref{fig:Q}, including the generation
of turbulence in the annular supersonic shear layer, the sudden break-up of the jet downstream and the radiation of intense waves in the
density field, known as Mach waves. These Mach waves are observed in the shadowgraphy images of~\citet{canchero16}
and corroborate the findings shown here.
Finer turbulent scales can be seen downstream the Mach disk, associated with the bending of the shock (see the discussion in section~\ref{sec:shear_machdisk}).

The isosurface of the Q-criterion also shows that the initial part of the shear layer is not dominated by the Kelvin-Helmotz instability with its coherent azimuthal rollers, as usually observed in incompressible mixing layers. In the present case, oblique modes are observed to dominate the initial part of the shear-layer, leading to small-scale three-dimensional structures. This change is due to the high value of the local convective Mach number ($M_c \approx 1.03$) at the beginning of the shear layer, in agreement with the findings of~\citet{sandham_reynolds_1991}, who demonstrated that when $M_c > 1$, the oblique modes dominate the instability process in planar compressible shear layers. Similar structures were observed by~\citet{deck2007} in the supersonic annular shear layer past an axi-symmetric trailing edge, characterised by a convective Mach number greater than one.


\section{Analysis of the wall-pressure signature}
\label{sec:wallp}

Characteristics of the wall pressure are described in this section. First, in 6.1, it is evaluated how the spectral
content of the wall-pressure fluctuations vary throughout the axial direction of the nozzle.
In 6.2 our view of the physics is enhanced by further considering azimuthal
decompositions of the fluctuating wall-pressure field.

\subsection{Evolution along the longitudinal axis}

\begin{figure}
 \centering
	\includegraphics[width = 0.97\textwidth]{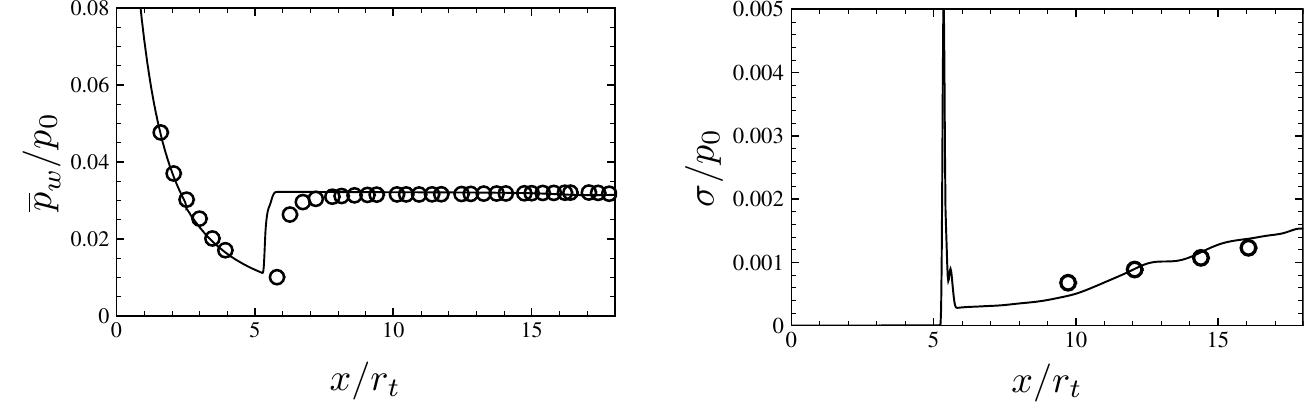}
   \caption{Distribution of (left) mean wall pressure and (right) standard deviation of wall pressure fluctuations.
	    Solid line, DDES; open circles, reference experimental data.}
   \label{fig:pwall}
\end{figure}

The distribution of the mean wall pressure ($\overline{p}_w$), averaged in time and in the azimuthal direction, is shown in figure~\ref{fig:pwall}a compared with the experimental data obtained from the static sensors. The wall pressure decreases until the incipient separation point, then the oblique shock causes an abrupt increase up to a plateau value close to the ambient pressure level. 
The numerical data show a very good agreement with the experiments, and a small difference is only observed in the average position of the separation point, that can be attributed to the unavoidable delay in the transition from the RANS to the LES mode~\citep{Shur2015}.  We point out that the remarkable agreement between DDES and the experiment in the prediction of the plateau pressure value is a strong evidence that the LES mode is fully active in the subsonic recirculation zone and that the simulation correctly captures the exchange of momentum between the main jet and the separated flow, dominated by the high convective Mach number of the shear layer~\citep{pantano02}.

The standard deviation of the wall pressure signals ($\sigma$) is shown in figure~\ref{fig:pwall}b,
together with the corresponding experimental values obtained from the dynamic sensors.  
The behaviour of $\sigma$ along the x-axis is typical of shock-wave/turbulent boundary layer interactions~\citep{Dolling1985}, characterised by a dominant sharp peak at the separation-shock location, followed by a lower level downstream, where the turbulent shear layer develops.
The experimental data points are located well downstream of the shock-separation region and match well with the numerical data, except at the first experimental point, probably due to the delay in the formation of the turbulent shear layer in the numerical simulation. 
\begin{figure}
 \centering
	\includegraphics[width = 0.8\textwidth]{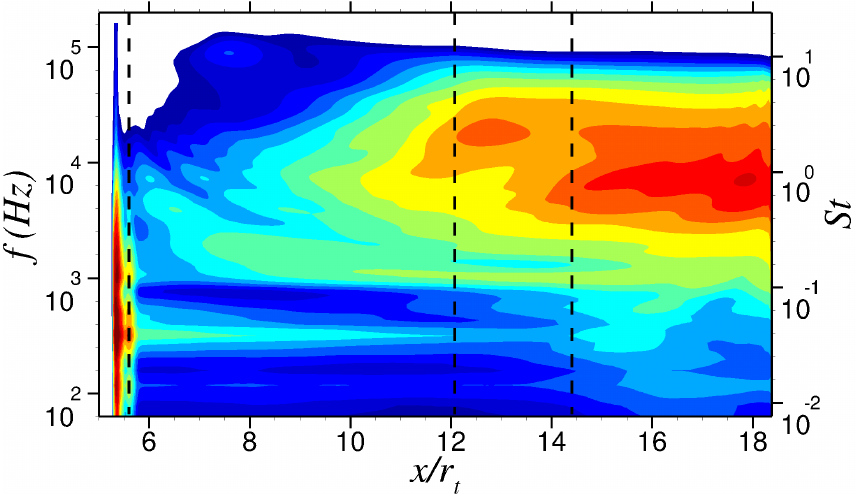}
   \caption{Contours of premultiplied power spectral densities of the wall-pressure signals as a function of the streamwise location and frequency.
	    Sixteen contour levels are shown in exponential scale between 0.0001 and 1.}
    \label{fig:map_spec}
\end{figure} 

The premultiplied wall-pressure spectra $f \, G(f)$ are reported in
figure~\ref{fig:map_spec} as a function of both the longitudinal coordinate $x$ and frequency $f$.
This representation as a two-dimensional map provides a complete picture of the spatial distribution
of the energy along the nozzle wall and of the contribution of the different frequencies to the total energy of the signal.
The power spectral densities (PSD) have been estimated based on the Welch method, i.e., subdividing the overall
pressure record into 12 segments with 50\% overlapping, which are individually Fourier-transformed.
The frequency spectra are then obtained by averaging the periodograms of the various segments, thus minimizing
the variance of the PSD estimator. Finally, for visualisation purposes, the Konno-Ohmachi smoothing~\citep{konno1998ground}
has been applied, whose filter function guarantees a constant bandwidth on a logarithmic scale.
The spectral map shows that there are two regions with high fluctuation energy, according to the distribution of the standard deviation.
The first zone is located in correspondence of the separation point ($x/r_t = 5.3$), where the signature of the shock motion is visible, characterised by a broad peak in the low-frequency range and a narrow footprint in the spatial direction.
The second region characterised by high levels of pressure fluctuations is located in the high-frequency range (around $f = 10^4$ Hz) and is associated with the development of the separated shear layer, whose convected vortical structures radiates pressure disturbances which increase in intensity downstream. 
It is worth to point out that, overall, the distribution of the pressure fluctuation energy shown in figure~\ref{fig:map_spec} is qualitatively similar to that previously observed in canonical shock wave boundary layer interactions~\citep{Dupont2006}, despite the significant differences in the geometrical configuration and shock topology (open separation bubble) of the present flow case.
\begin{figure}
 \centering
	\includegraphics[width = 0.97\textwidth]{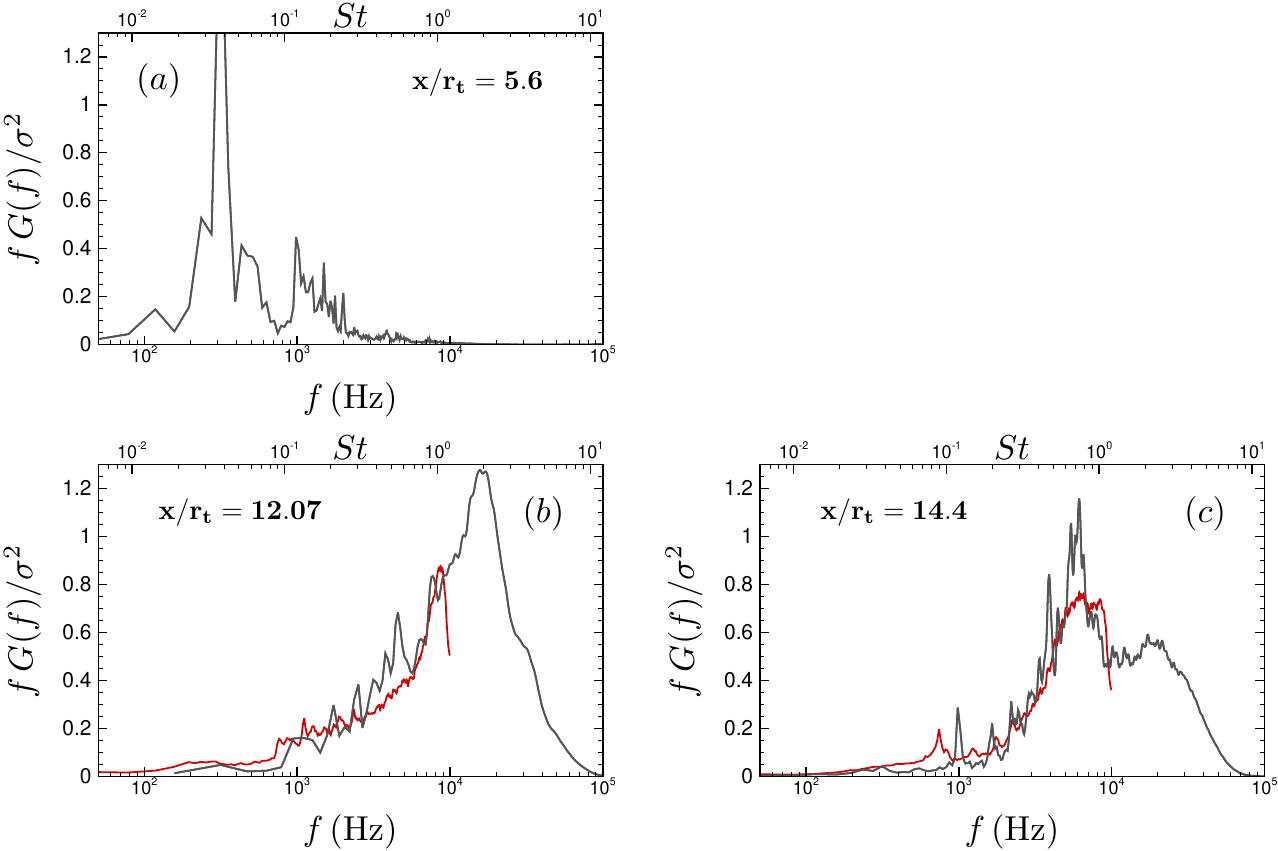}
  \caption{Normalized premultiplied power spectral densities of the wall pressure signals at different x-stations.
           Solid grey line, DDES; solid red line, experimental data.}
  \label{fig:specx}
\end{figure}

To better visualise the frequency content of the wall-pressure fluctuations, figure~\ref{fig:specx} shows the premultiplied spectra
at three representative axial stations corresponding to the vertical lines in figure~\ref{fig:map_spec}.
Note that, for the purpose of comparison, the curves are normalized by
the integral of the spectra over the range of frequencies
that are being resolved by the experiment.
The first station is selected immediately downstream of the separation point ($x/r_t = 5.6$), whereas the other two
correspond to positions where experimental data from dynamic probes are available ($x/r_t = 12.07$ and $x/r_t = 14.40$).
At $x/r_t = 5.6$, the spectrum corresponding to the shock motion shows the presence of a high-amplitude, broad bump,
between 100 Hz and 2000 Hz approximately, with a maximum at $f \approx 316 Hz$ and a secondary peak located at $f \approx 1000 Hz$.
According to~\citet{Baars2015} the low-frequency peak can be attributed to an acoustic resonance, see for example~\citet{wong:05jpp} and references therein, occurring between the exit plane and the shock system, described by the one-quarter standing wave model in a open-ended pipe.
The fundamental frequency predicted by the model is given by  
\begin{equation}
f_{ac} = \frac{a_{\infty}(1-M_{N.E.}^2)}{4 L},
\end{equation}
where $a_{\infty}$ is the ambient 
speed of sound, $M_{N.E.}$ is the Mach number in the separated region at the nozzle exit and $L$ is the distance between the averaged separation position and the nozzle lip. Using the data extracted from DDES, $a_{\infty} = 345$ m/s 
and $L = 0.247 m$, we obtain an acoustic frequency $f_{ac} = 340 Hz$, that well compares
with the peak in the spectrum at 316 Hz.
Moving downstream a different picture emerges, 
characterised by a shift of the spectra towards the high-frequency range
(around $f = 10^4 Hz$), caused by the three-dimensional turbulent structures of the developing shear layer.
As expected, the peak frequency is observed to shift at lower values moving in the streamwise direction,
as a consequence of the growing turbulent eddies. Interestingly, a peak located at intermediate frequencies
($f = 992 Hz$ for DDES, $f = 742 Hz$ for the experiment) is visible in the spectra downstream of the shock location,
whose nature will be investigated in the following.
The scenario described above is depicted by both the numerical and experimental spectra, which are found to display
a very good agreement except very close to the maximum frequency resolved by the measurements,
where the rapid roll-off of the spectra is most likely due to filtering effects of pressure sensors.
It is worth to highlight that the main features of the power spectral densities here described can also be found in previous experimental studies performed for various TOP nozzles at NPR's corresponding to the FSS pattern~\citep{Nguyen2003, Baars2012,Verma2014} and a different TIC nozzle~\citet{Jaunet2017}.
To better characterise the intermediate peak it is useful to introduce a Strouhal number, defined as~\citep{Tam1986,canchero16}:
\begin{equation}
St = f \frac{D_j}{U_j},
\end{equation}
where $U_j$ is the fully expanded jet velocity
\begin{equation}
U_j = \sqrt{\gamma R T_0}\frac{M_j}{\sqrt{1+\frac{\gamma-1}{2}M_j^2}},
\end{equation}
$T_0$ is the stagnation temperature, $\gamma$ the specific heat ratio, $R$ the 
air constant, $M_j$ is the fully adapted Mach number, which is a function of the nozzle pressure ratio through the isentropic relation
\begin{equation}
\frac{p_0}{p}=\Bigg(1+\frac{\gamma-1}{2}M_j^2\Bigg)^{\frac{\gamma}{(\gamma-1)}}.
\end{equation}
The length-scale $D_j$ is computed as a function of $M_j$, the design Mach number $M_d$ and the nozzle exit diameter $D$ through the mass flux conservation
\begin{equation}
\frac{D_j}{D}=\Bigg(\frac{1+\frac{\gamma-1}{2}M_j^2}
{1+\frac{\gamma-1}{2}M_d^2}\Bigg)^{\frac{\gamma+1}{4(\gamma-1)}} \Big(\frac{M_d}{M_j}\Big)^2.
\end{equation}
\begin{figure}
 \centering
	\includegraphics[width = 0.8\textwidth]{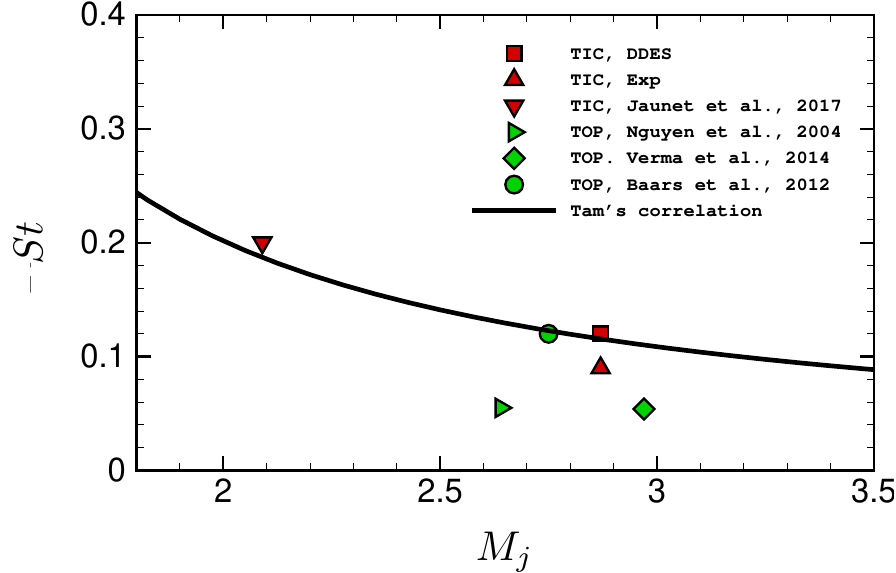}
   \caption{Strouhal number of the intermediate peak in the frequency
spectrum at the shock position as a function of the fully adapted Mach number $M_j$ for different geometries and NPR's.}
    \label{fig:st}
\end{figure}

The non-dimensional frequencies of the intermediate peak from the different experimental flow cases cited above have been reported in figure~\ref{fig:st} as a function of $M_j$. A decreasing trend of $S_t$ with the fully adapted Mach number seems to emerge from the data. This behaviour was also identified by~\citet{Jaunet2017}, who proposed that this peak could be 
attributed to a sort of screech~\citep{Raman99},
an aeroacoustic feedback mechanism involving downstream propagating disturbances in the shear layer, 
the interaction with the shock cells and upstream propagating acoustic waves.
According to expression proposed by~\citet{Tam1986} the screech frequency $f_{sc}$ can be evaluated as
\begin{equation}
	St_{sc} = f_{sc}\frac{D_j}{U_j}=0.67(M_j^2-1)^{-1/2}\Big[1+0.7M_J(1+\frac{\gamma-1}{2}M_j^2)^{-1/2}\Big(\frac{T_a}{T_0}\Big)^{-1/2}\Big] 
\end{equation}
where $T_a$ is the ambient temperature.
This correlation, plotted in figure~\ref{fig:st}, shows that the characteristic screech frequency decreases with increasing $M_j$ (or equivalently with increasing nozzle pressure ratio), mainly due to the shortening of the shock cells. 
Despite a certain level of dispersion, the proximity of the experimental data and of the DDES prediction to the empirical correlation suggests that the intermediate peak could be attributed to a screech phenomenon. 

\begin{figure}
\centering
	\includegraphics[width = 0.97\textwidth]{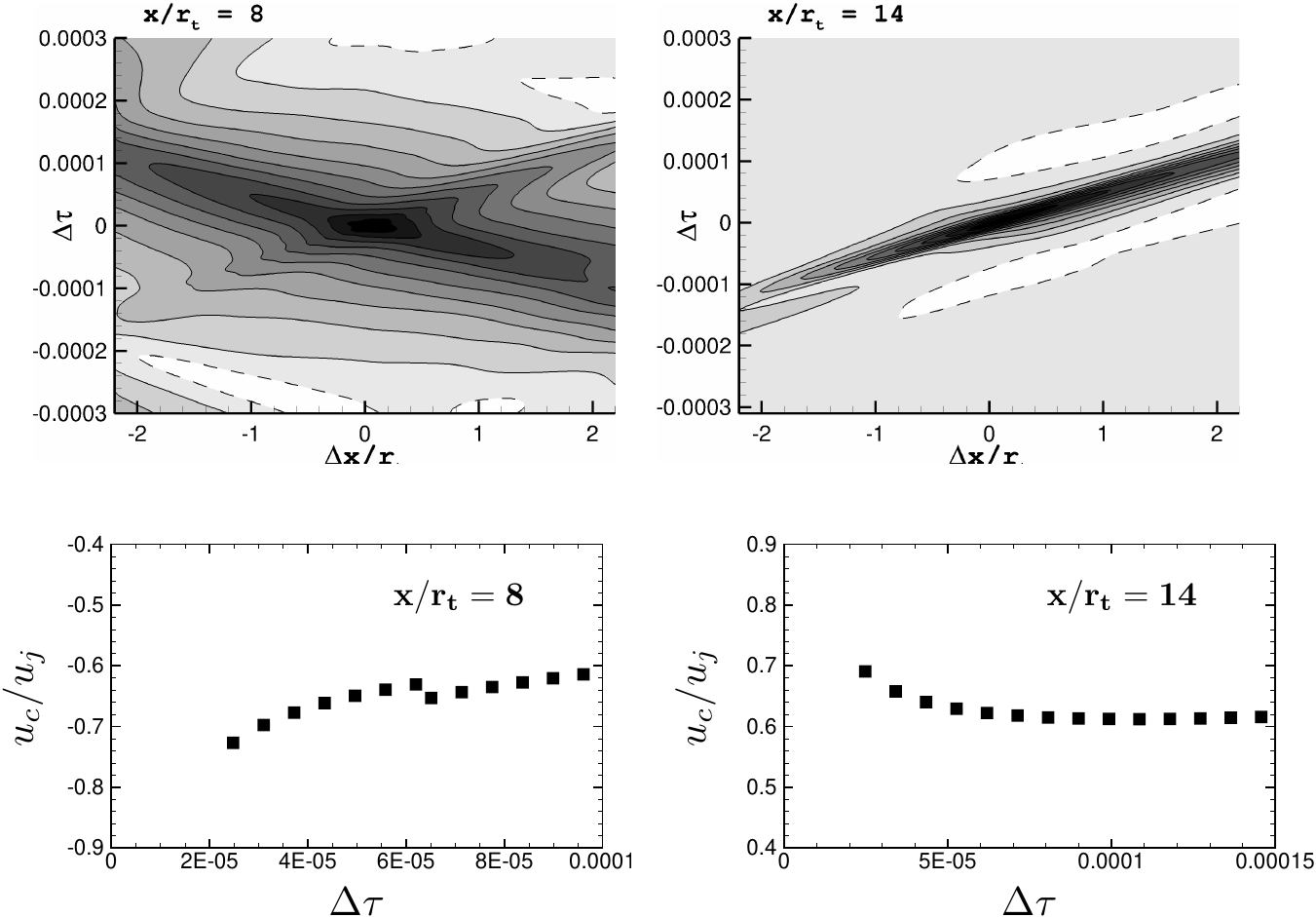}
 \caption{Top: contours of the space-time pressure correlation coefficient $C_{pp}(\Delta x,0,\Delta \tau)$. Eleven contour levels are shown in 
	  the range $-0.1<C_{pp}<0.9$. Bottom: local convection velocity of wall pressure fluctuations as a function of the time separation $\Delta \tau$.}
 \label{fig:corr}
\end{figure}

Additional insights into the propagation of pressure disturbances, in terms of direction and velocity, can be gained by inspection of the space-time correlation coefficient, defined as
\begin{equation}
	C_{pp}(x,\Delta x,\Delta \theta,\Delta \tau)= \frac{R_{pp}(x,\Delta x,\Delta \theta,\Delta \tau)}{\left[ R_{pp}(x,0,0,0)\right]^{1/2} \, \left[ R_{pp}(x,\Delta x,0,0)\right]^{1/2}},
\end{equation}
where
\begin{equation}
	R_{pp}(x,\Delta x,\Delta \theta,\Delta \tau) = \overline{p_w^{'}(x,\theta,t) \, p_w^{'}(x+\Delta x,\theta+\Delta \theta,t+\Delta \tau)}
\end{equation}
is the space-time correlation function, $\Delta x$ and $\Delta \theta$ are the spatial separations in the streamwise
and azimuthal directions, $\Delta \tau$ is the time delay,
and the overbar denotes averages taken with respect to the
azimuthal direction (exploiting homogeneity) and time. 
Figure~\ref{fig:corr} reports contours of the space-time correlation coefficient $C_{pp}(x,\Delta x,0,\Delta \tau)$ taken at two stations along the nozzle wall, 
$x/r_t= 8$ and $x/r_t = 14$.
In both cases the shape of the contours reflects the convective nature of the pressure field, but while at the latter station ($x/r_t = 14$) the pressure signal is mainly characterised by a coherent downstream propagation of pressure-carrying eddies, at the first location ($x/r_t = 8$) it is also well visible the presence of upstream-propagating disturbances, coming from the interaction of the vortex structures with 
the reflected shock and with the second Mach disk, located near the nozzle exit section.

From the maps of the space-time correlation coefficient it is possible to evaluate the convection velocities of the pressure-carrying eddies.  
Following~\citet{bernardini2011}, the convection speed
corresponding to a given time delay $\Delta \tau$ is defined as the ratio
$\Delta x/\Delta \tau$ taken at the spatial separation value $\Delta x$ where a local maximum of $C_{pp}$ is attained. The resulting convection speeds
are displayed in figure~\ref{fig:corr} (bottom panels), as a function of 
the time separation $\Delta \tau$. The convection velocities, 
in both directions, are observed to slightly decrease (in magnitude) with $\Delta \tau$, indicating that large-scale disturbances
move slower than the small eddies.
We point out that the convection speed of pressure disturbances is in the range between $0.6 u_j$ and $0.7 u_j$,
in agreement with the empirical value quoted by~\citet{Tam1986}.

\subsection{Azimuthal decomposition of the pressure field}
\label{azimuthal}

\begin{figure}
 \centering
	\includegraphics[width = 0.97\textwidth]{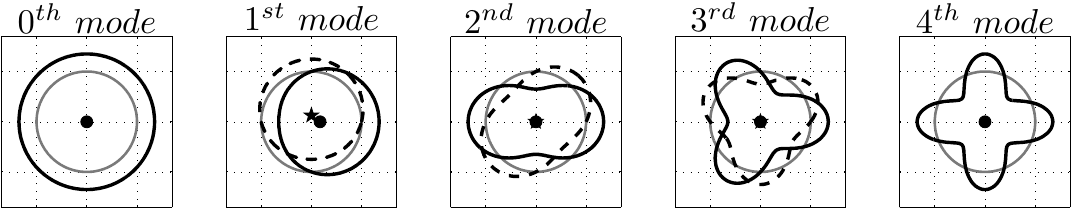}
 \caption{Physical interpretation of the Fourier azimuthal decomposition.
	  Solid and dashed lines correspond to the real ($m>0$) and imaginary ($m<0$) parts of the Fourier coefficients.}
 \label{fig:plot_fouriermodes}
\end{figure}
\begin{figure}
 \centering
	\includegraphics[width = 0.97\textwidth]{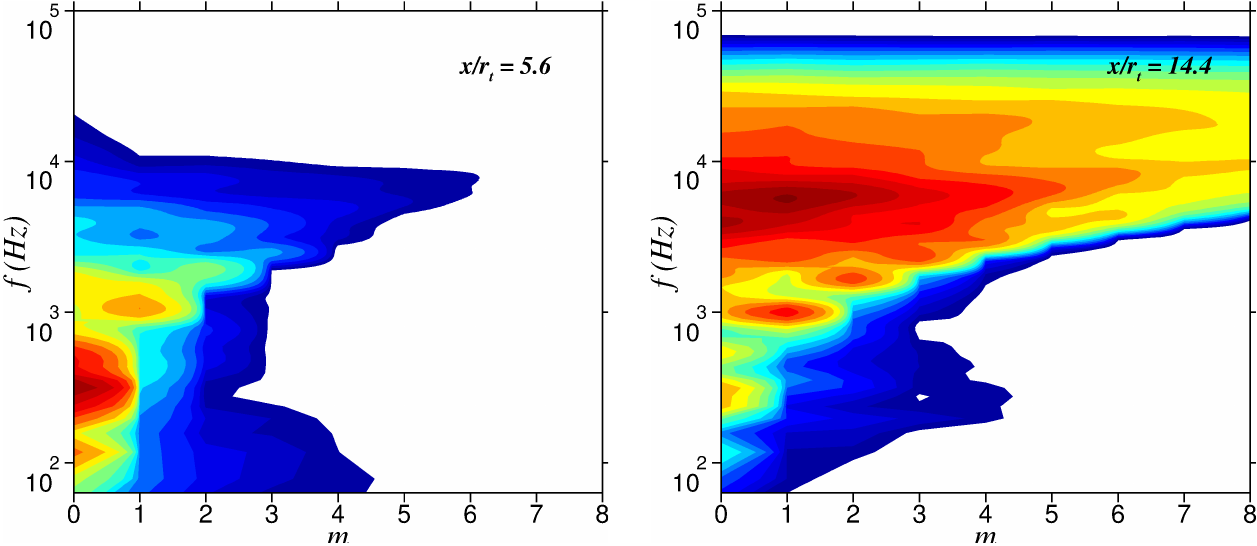}
  \caption{Contours of the premultiplied azimuthal wavenumber-frequency spectrum $f \phi_{pp}$ at two different axial stations.
	   Twenty contour levels are shown in exponential scale between 0.005 and 2.}
  \label{fig:wave_512_856}
\end{figure}
\begin{figure}
 \centering
	\includegraphics[width = 0.97\textwidth]{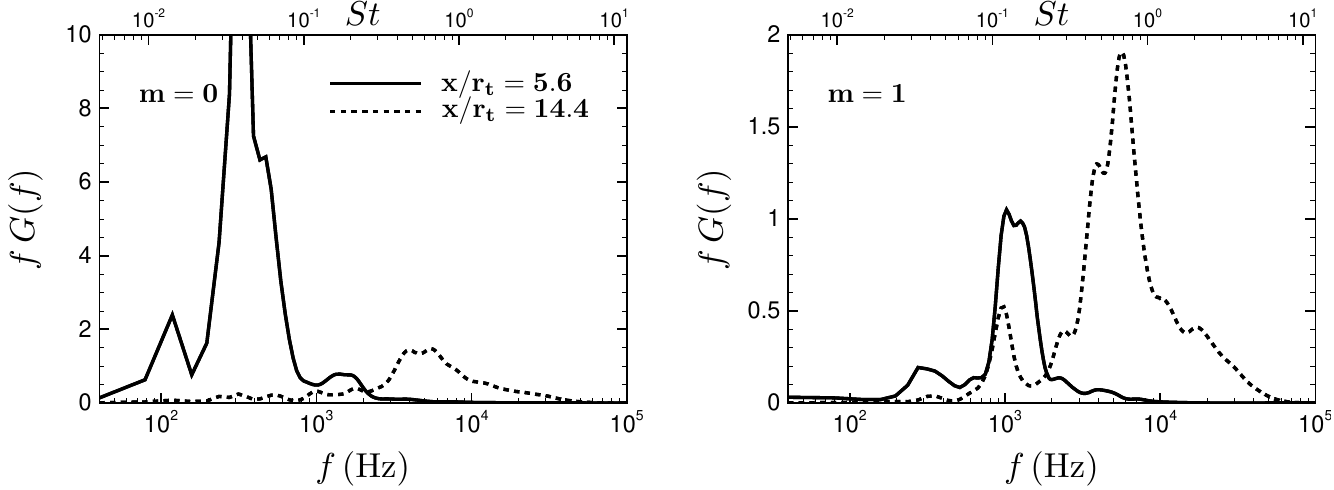}
 \caption{Premultiplied spectrum of the $0^{th}$ azimuthal Fourier mode (left) and of the $1^{st}$ azimuthal Fourier mode (right) 
          at two axial stations.}
 \label{fig:psdm0m1}
\end{figure}

To gain insights into the origin of the aerodynamic loads and characterise how the energy is distributed among the different modes it is useful to carry out a Fourier-azimuthal decomposition of the unsteady wall pressure signature. 
To that purpose we consider the azimuthal wavenumber-frequency spectrum, defined as
\begin{equation}
	\phi_{pp} (x,m,f) = \int_{-\infty}^{\infty} \int_0^{2 \pi} R_{pp} (x,0,\Delta \theta, \Delta \tau) \, e^{-i (m \Delta \theta + f \Delta \tau)} \mathrm{d}(\Delta \theta) \, \mathrm{d} (\Delta \tau),
\end{equation}
where $m$ is the mode number in the azimuthal direction. We remind that, by symmetry considerations, only the antisymmetric mode $m = 1$ can provide a contribution to the aerodynamics side loads, as illustrated by figure~\ref{fig:plot_fouriermodes}, where a schematic of the first four azimuthal pressure modes is reported.
The wavenumber-frequency spectrum $\phi_{pp}$ is reported in figure ~\ref{fig:wave_512_856} for two representative stations, corresponding to a location close to the separation shock ($x/r_t = 5.6$) and in the recirculation bubble ($x/r_t = 14.4$).
In addition, to better quantify the evolution of the modes along the nozzle wall, single cuts are taken from the wavenumber-frequency maps for the first two azimuthal modes and are reported in figure~\ref{fig:psdm0m1}.
The figures reveal that the shock motion has a clear organisation in the azimuthal direction, with the dominant peak at $f = 316$ Hz ($St = 0.038$) exclusively associated with the axi-symmetric (breathing) mode $m=0$, and the secondary peak at $f= 1000$ Hz ($St = 0.12$) linked to the first ($m = 1$) Fourier mode. Essentially, the emerging picture is that of a shock characterised by a low-frequency piston-like motion, with the excitation of non symmetrical modes at intermediate frequency.
Moving at the downstream station ($x/r_t = 14.4$) the contribution of the breathing mode at low frequencies is strongly reduced, 
whereas the two peaks located at $f = 1000$ Hz ($St = 0.12$), $m = 1$ still persist. 
As previously observed in the spectra of figure~\ref{fig:map_spec}, at this station most of the energy is contained in the high-frequency range, and it is spread in a wide range of azimuthal modes, reflecting the turbulent character of the wall-pressure fluctuations in the recirculation region.

\section{Annular supersonic shear layer and Mach disk region}
\label{sec:shear_machdisk}
\subsection{Coherent structures dynamics and spectral analysis}

Instantaneous flowfields are used in this section to assess the existence
and evolution of coherent vortices along the detached shear layer.
\begin{figure}
 \centering
	\includegraphics[width = 0.97\textwidth]{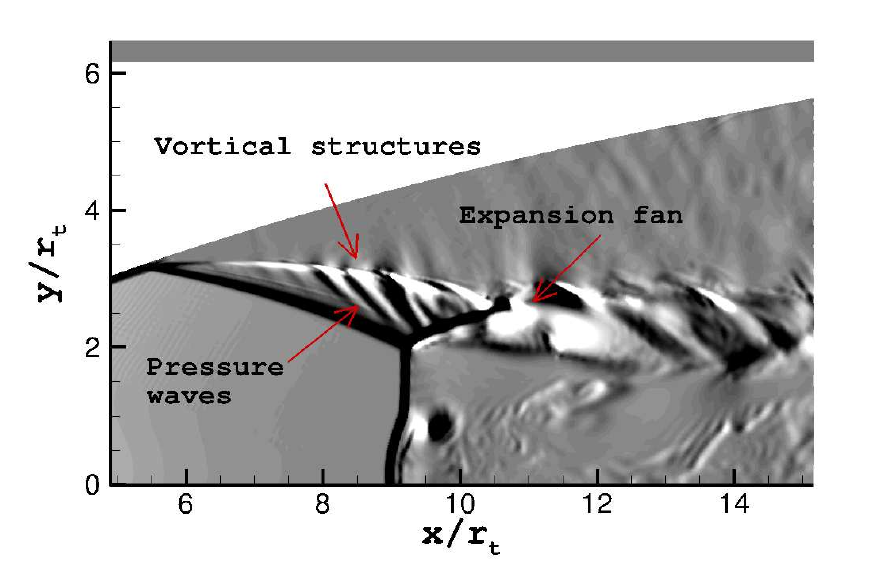}
 \caption{Visualization of pressure waves radiated from vortical structures of the separated shear layers,
	  educed using contours of $\partial \rho/\partial x$.}
 \label{fig:shear}
\end{figure}
Figure~\ref{fig:shear} shows the streamwise derivative of the density in the first part of the separated shear layer. 
The black areas indicate compression (or shock) regions, while the white areas indicate expansion zones.  
It is apparent from the figure the formation of coherent vortices at the beginning of the shear layer,  highlighted by the presence of pressure waves radiating in the lower side (where the flow is supersonic) that interacts with both the separation and the reflected shock.
The reflection of the latter shock as an expansion fan when it interacts with the upper boundary of the annular shear layer is also well visible in the figure. 
\begin{figure}
	 \centering
	  \includegraphics[width = 0.97\textwidth]{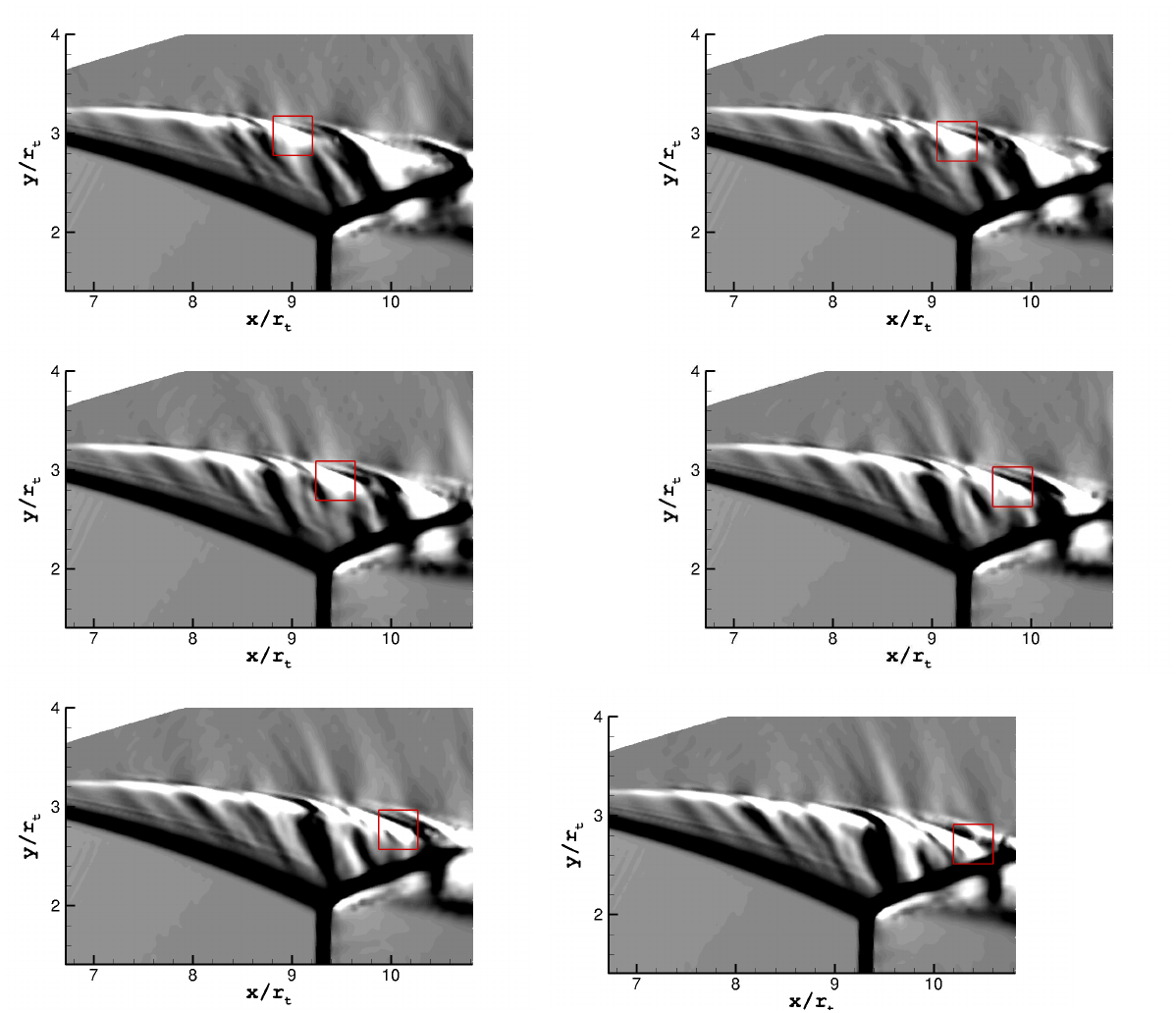}
	   \caption{Convection of coherent structures in the separated shear layer, shown by means of contours of $\partial \rho/\partial x$. 
            Instantaneous snapshots $\Delta t = 1.3 \times 10^{-5}$ s apart from left to right.}
	    \label{fig:prova}
\end{figure}
The evolution of vortical structures in the shear-layer is tracked in figure~\ref{fig:prova}, where a sequence of snapshots of the streamwise component of the density gradient is reported. 
The eddies appear to be characterized by an elliptical shape and to be convected without significant changes of their shape, and without merging, until they interact with the reflected shock.
It is worth to point out that our findings are very similar to those reported by~\citet{deck2007} for the analysis of the vortex dynamics in the compressible shear layer past an axi-symmetric trailing edge: in that case some merging between vortices was visible, but without the well-known rotational pairing which contributes to the growth rate of subsonic shear layers.  

\begin{figure}
 \centering
	\includegraphics[width = 0.7\textwidth]{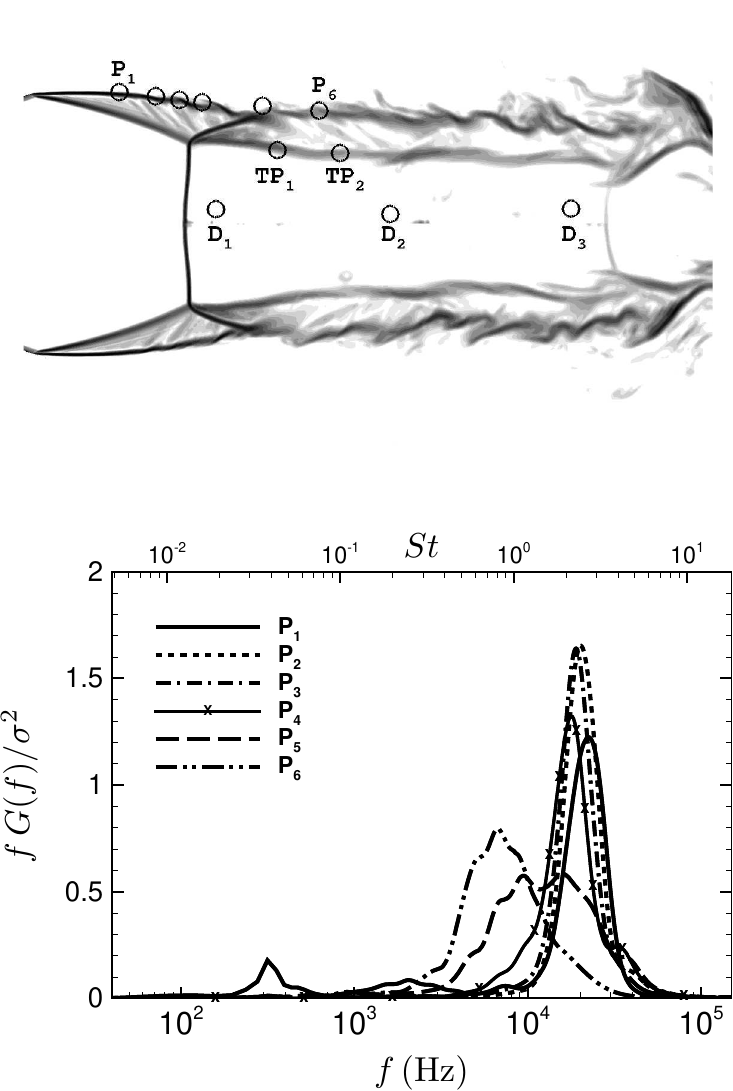}
   \caption{Top) position of the numerical pressure probes; bottom) pressure spectra along the separated shear layer (from $P_1$ to $P_6$).} 
	    \label{fig:psd_field}
\end{figure}

To complete the description of the separated shear layer, figure~\ref{fig:psd_field} shows the
results of a spectral analysis of the pressure fluctuations, conducted considering
six probes, denoted as $P_{1-6}$.
The first four probes are located before the interaction of the reflected shock with the shear layer, 
whereas the last two pressure sensors are placed downstream of this interaction. The signal from  sensor $P_1$ is characterised by a first small peak centred around 316 Hz ($St = 0.038$), that is connected to the shock movement. The 
second high-frequency and high-energy broad peak is centred around 20 kHz ($S_t=2.41$) 
and it is the signature of the instability process of the shear layer. 
Moving downstream, the high-frequency component spreads to lower frequencies and
the peaks of the spectra broaden, reflecting the growth of the turbulent structures in the developing shear layer.
The contribution of the low-frequency peak rapidly disappears downstream, compared to the high-energy content of the pressure fluctuations given by the turbulent structures. 

\begin{figure}
	 \centering
	  \includegraphics[width = 0.97\textwidth]{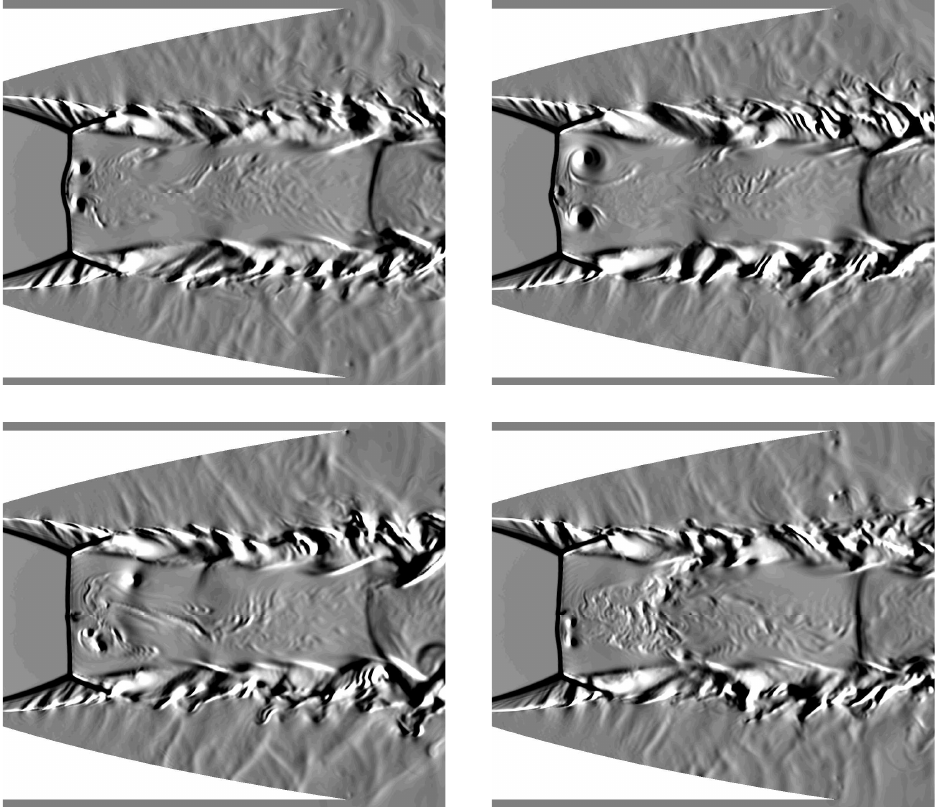}
	   \caption{Convection of coherent structures downstream of the Mach disk, shown through contours of $\partial \rho/\partial x$. 
		    Instantaneous snapshots at $t$ = $3.36 \times 10^{-2}$ s, $t$ = $3.39 \times 10^{-2}$ s, $t$ = $3.43\times 10^{-2}$ and $t$ = $3.46\times 10^{-2}$ s from 
		    left to right and top to bottom. }
	    \label{fig:mach_disk}
\end{figure}

The region downstream of the Mach disk is characterised by an intense vortex shedding activity and consequent production of vorticity, well highlighted by the sequence of snapshots of $\partial \rho/\partial x$ reported in figure~\ref{fig:mach_disk}. 
In the first frame, the Mach disk is bent upstream and two counter-rotating vortices are emitted. The bending of the Mach disk is caused by two factors as discussed in~\citet{Nasuti2009}. First of all, the Mach number field upstream of the shock region 
is not uniform also in the radial direction, as clearly shown in figure~\ref{fig:mach}.
This means that, to adapt to the value of the static pressure in the subsonic flow 
(which is nearly uniform in a time and space average) the shock has to change its curvature in the radial direction in order to change its intensity. Secondly, the flow unsteadiness 
causes a continuous variation of the static pressure behind the shock, inducing the oscillation of the Mach disk and the variation of its curvature. The variation of the 
shock intensity in the radial direction causes an entropy gradient downstream and, as described by the Crocco's theorem~\citep{Crocco}, vorticity is produced.
These structures are then convected downstream and impinge on the annular supersonic shear layer and on the second shock cell, in a non-symmetric way. The last picture (at $t = 3.46\times 10^{-2}$ s) shows the formation of two new counter-rotating vortices, but this time they are located below the centre line of the nozzle, indicating again the 
non-symmetric behaviour of this shedding phenomenon.
An inspection of the flow field along the simulation time also reveals that this activity is not continuous 
but rather intermittent, as clearly highlighted by the accompanying movie
available online.

To further investigate the flow dynamics, several pressure probes have been placed downstream of the Mach disk (near the nozzle axis) and in the inner shear layer which originates from the triple point of the separation shock reflection, as shown by figure~\ref{fig:psd_field}.
\begin{figure}
 \centering
	\includegraphics[width = 0.97\textwidth]{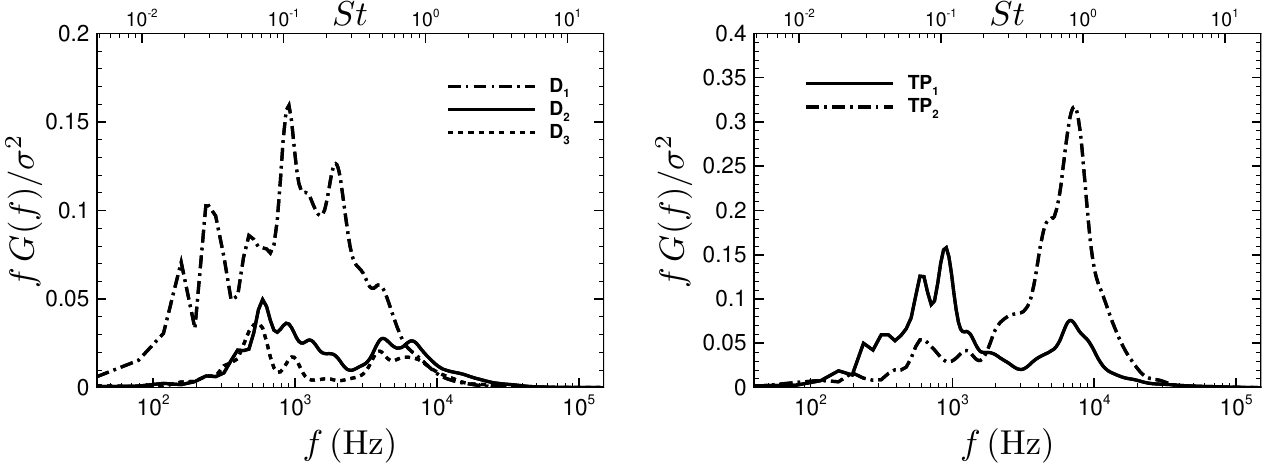}
 \caption{Premultiplied frequency spectra of the pressure signal downstream of the Mach disk (from $D_1$ to $D_3$)
	and along the internal shear layer past the triple point ($TP_1$ and $TP_2$).} 
    \label{fig:psd_md}
\end{figure}
Figure~\ref{fig:psd_md} (left panel) shows the premultiplied spectra downstream of the Mach disk. 
The signal from the first probe ($D_1$) is characterised by a very broad bump, with the highest peak at 905 Hz ($St = 0.11$). This fluctuation energy is originated by the shock unsteadiness and its amplitude decreases moving downstream, as shown by the spectra at $D_2$ and $D_3$. 
The internal vortices shed by the Mach disk irradiate pressure waves as they are 
convected, stretched and broken in smaller scales. This turbulence noise explain the broad shape of the spectra. 
The peak at 905 Hz ($St = 0.11$) dominates also the initial part of the inner shear layer, as shown by the spectrum from the 
probe $TP_1$ in the right panel of figure~\ref{fig:psd_md}. This frequency is very close,
and most likely correlated, to the intermediate peak frequency (1000 Hz, $St = 0.12$) found in the wall pressure signature and attributed to 
the first Fourier azimuthal mode (see the discussion in Sec.~\ref{sec:wallp}).
Again, moving from $TP_1$ to $TP_2$, the amplitude of this peak decreases while the contribution from the developing turbulent shear layer becomes more and more important and it is characterised by a broad peak around 7000 Hz ($St = 0.843$).

\subsection{Discussion of the feedback mechanism}

\begin{figure}
 \centering
	\includegraphics[width = 0.7\textwidth]{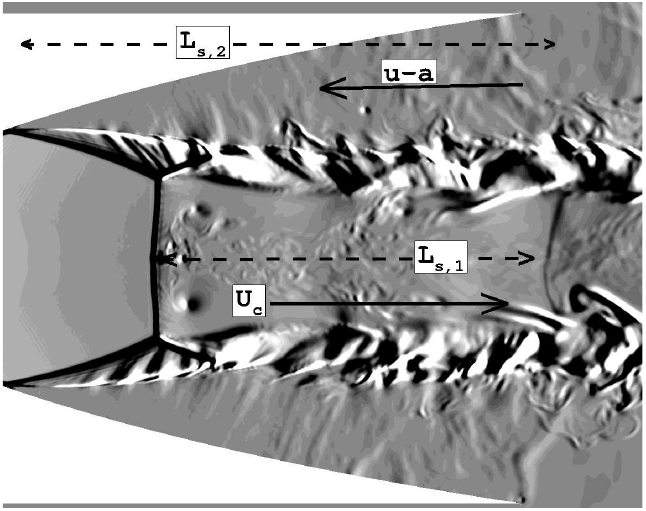}
   \caption{Visualization of the proposed feedback loop in the shock cell within the nozzle.} 
    \label{fig:feed}
\end{figure}

It has been shown that the peak at $f = 1000$ Hz ($St = 0.12$) agrees with
the correlation proposed by Tam and that it is associated with the first Fourier azimuthal mode ($m = 1$). 
A similar agreement was also displayed by the experimental data of~\citet{Jaunet2017}. Those authors
did not find any trace of the screech tone in the external ambient and speculated the existence of
screech-like mechanism inside the nozzle, sustained by the presence of the internal subsonic flow region
downstream of the Mach disk, which may provide a support for possibly upstream propagating waves.
The numerical findings of the present DDES seem to provide support to the existence of an internal
feedback loop.
Following the model proposed by~\citet{Powell53}, which states 
that the screech temporal period can be considered as the sum of the time taken by the flow disturbances
to propagate downstream one shock cell and the time needed by the acoustic waves outside the jet to propagate
back over the same distance towards the nozzle lip, we propose a path for the feedback loop as shown in figure~\ref{fig:feed}. 
The vortical structures coming from the Mach disk and the triple point have to cover the distance $L_{s,1}=$ 0.194 m before interacting with the 
second shock, whose perturbation causes the emission of acoustic waves. 
Since the flow upstream of the second shock is supersonic, the acoustic disturbances 
can travel upstream, with velocity $u-a$, 
only through the turbulent recirculating zone external to the annular supersonic shear 
layer and adjacent to the nozzle wall. This path involves the 
length $L_{s,2}=$ 0.26 m between the second shock and the separation line. 
When these acoustic waves reach the separation line, the whole shock system is perturbed and 
new instability waves are emitted from the detached shear layer and from the Mach disk.
Therefore, as a first approximation, the total period of the loop, $T_{sc}$, can be modelled
taking into account the distance $L_{s,1}$ for the convective downstream movement and the distance $L_{s,2}$ for the acoustic upstream movement:
$$T_{sc} = \frac{L_{s,1}}{u_c} + \frac{L_{s,2}}{|u-a|}$$
giving a screech frequency $f_{sc}$:
$$f_{sc} = \frac{u_c}{L_{s,1}+L_{s,2}\frac{u_c}{|u-a|}}$$
Considering, approximately, a speed of sound in the subsonic separated region $a = 345$ m/s, 
a longitudinal velocity $u = -80$ m/s and a 
convective velocity $u_c \approx 0.65 u_j$, 
we found $f_{sc} = 910$ Hz ($St = 0.11$), a value reasonably close to the numerical
prediction extracted from the spectra ($f = 1000$Hz, $St = 0.12$).

\section{Conclusions}
\label{sec:conclusions}

We have performed a delayed detached eddy simulation of an overexpanded separated flow in a TIC nozzle featuring a free shock
separation (FSS), with the aim of characterising the unsteady pressure loads caused
by the self-sustained shock motion and identifying the driving physical mechanisms.
The DDES approach revealed itself to be a very powerful tool to investigate the flow dynamics
at conditions (Reynolds number $Re = 10^7$, NPR = 30.35) for which the shock system causes the
flow separation well inside the nozzle, in a region that is not easily accessible to experimental instruments.
The simulation results were validated against a comparison with measurements performed at the University of Texas at Austin,
and a very good agreement was obtained in the distribution of mean and fluctuating wall-pressure statistics. 

Fourier analysis in time and in the azimuthal direction has been carried out to describe the wall-pressure signature.
We found that the spectrum at the separation point is characterised by a peak at $f = 316$ Hz ($St = 0.038$) and azimuthal mode $m = 0$,
implying a low-frequency breathing (piston-like) motion of the shock system.
This symmetric mode is most likely driven by an acoustic resonance, since its main frequency is well
predicted by the one-quarter standing wave model. A secondary peak at intermediate frequencies ($f \approx 1000$ Hz, $St = 0.12$)
was also observed, associated in the wavenumber space with the first azimuthal mode ($m = 1$), which is identified as the cause
of the aerodynamic side loads. This non-symmetric pressure mode is also found to persist at the same frequency
in the spectrum of the wall-pressure fluctuations in the turbulent recirculation zone.
The characteristic frequency ($f \approx 1000$ Hz) of this phenomenon reasonably agrees
with the empirical correlation proposed by~\citet{Tam1986} for the screech, suggesting the presence
of a feedback loop similar to that observed for overexpanded external jets.
Motivated by this observation a feedback-loop model has been proposed to identify the driving physical mechanisms
in nozzle flows characterized by the free-shock separation pattern. 
The loop starts with the turbulent structures of the detached shear layer and with the intermittent vortex shedding activity
of the Mach disk, occurring at a frequency comparable with the peak frequency of the wall pressure at the first azimuthal mode.
These vortices propagate in the streamwise direction and interact
with the second shock cell, irradiating acoustic waves able to travel upstream in the outer subsonic region.
Finally, these perturbations excite new instabilities of the shear layer and of the shock system, closing the loop.
These results thus provide support to the speculation recently made by~\citet{Jaunet2017} on the importance
of the internal subsonic region in the feedback loop and, most importantly, underline the key role
of the strong vortex shedding activity of the Mach disk.

\begin{acknowledgments}
M. Bernardini was supported by the Scientific Independence of Young Researchers program 2014 (Active Control of Shock-Wave/ Boundary-Layer Interactions project, grant RBSI14TKWU), which is funded by the Ministero Istruzione Universit\`a e Ricerca. The simulations have been performed thanks to computational resources provided by the Cineca Italian Computing Center under the Italian Super Computing Resource Allocation initiative (ISCRA B/DDES-TIC/HP10BZR88R). 
\end{acknowledgments}
\bibliographystyle{jfm}
\bibliography{biblio}

\end{document}